\documentclass[aps,pra,showpacs,groupedaddress,twocolumn]{revtex4}
\usepackage{amsfonts}
\usepackage{graphicx}
\usepackage{latexsym}
\usepackage{makeidx}
\usepackage{amsmath}
\usepackage{amssymb}

\begin{document}

\title{Continuous intracavity monitoring of the dynamical Casimir effect}
\author{A V Dodonov}
\affiliation{Instituto de F\'{\i}sica and International Center for Condensed Matter Physics, Universidade de Bras\'{\i}lia, PO Box 04455,
70910-900, Bras\'{\i}lia, DF, Brazil \\
ITAMP, Harvard--Smithsonian Center for Astrophysics, Cambridge, MA 02138, USA}

\begin{abstract}
Dynamical Casimir effect (DCE) is the name assigned to the process of
generating quanta from vacuum due to an accelerated motion of macroscopic
neutral bodies (mirrors) or time-modulation of cavity material properties,
as well as the simulation of such processes. Here I review the
theoretical results on the detection of DCE using intracavity quantum
detectors, such as multi-level atoms, atomic networks and
harmonic oscillators. I also stress the mathematical equivalence of this problem
to the physics of optical parametric oscillators interacting with atoms or quantum wells, studied in Quantum Nonlinear Optics.
\end{abstract}

\maketitle


\section{Introduction}\label{sec1}

Dynamical Casimir effect (DCE) is the term used nowadays for a rather wide
group of phenomena whose common feature is the creation of quanta from the
initial vacuum state of some field due to time-modulation of material
properties or boundary conditions of some macroscopic system. In the
majority of cases considered so far this corresponds to the creation of
photon pairs from the electromagnetic vacuum due to the motion of a
mirror in empty space, oscillation of a cavity wall or modulation
of the dielectric properties of the medium inside it. Non-electromagnetic analogs of the DCE are also possible, such
as the recent report on generating correlated pairs of elementary excitations from the initial thermal state in a trapped Bose-Einstein
condensate (BEC) \cite{bec}, where phonons were created instead of photons. For short reviews on the history and recent achievements
in the DCE research see \cite{vdodonov,nori-review} and
references therein.

The detection, characterization and, ultimately, manipulation of excitations
created due to the DCE are as important as the very generation from vacuum.
For photons generated in empty space or in cavities one can
distinguish at least four different measurement schemes. In the \emph{%
extracavity detection} the photons propagate away from the source (mirror or
leaky cavity) and are ultimately detected using standard optical or
microwave techniques. This approach was discussed for photon emission from a
semiconductor microcavity \cite{liberato7} and superconducting resonator
\cite{liberato9} due to time-modulation of the Rabi frequency. For
superconducting coplanar waveguides terminated by a quantum interference
device (SQUID) the measurement setups were discussed in \cite%
{nori-prl,nori-pra,nori-review, nori-arx}, and the experimental detection of
radiation produced due to an analog of a moving mirror was described in \cite%
{nori-n}. Preliminary results on the detection of DCE photons produced in a lossy resonator composed of Josephson metamaterial were described in \cite{hakonen}.

\vspace{1cm}
\begin{figure}[th]
\includegraphics[bb = 0 0 200 180]{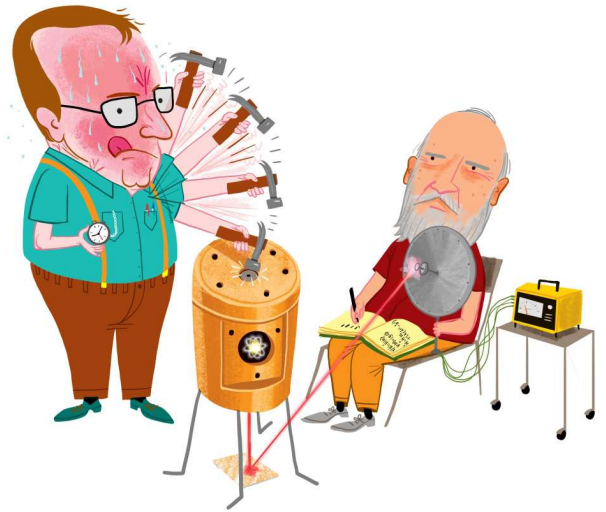}
\caption{Artistic view of the continuous intracavity monitoring of DCE
(illustration by \emph{Porf\'{\i}rio}). A slow unexcited atom (quantum detector)
crosses the cavity whose frequency is time-modulated by an experimentalist
who hits its wall at a constant rate. The photon
generation from vacuum is monitored by continuously measuring the atomic
state using some external measurement device (in the picture the atomic
spontaneous emissions are registered by an antenna). Reproduced from
http://iopscience.iop.org/0953-4075/labtalk-article/48481.}
\label{f0}
\end{figure}

For DCE implemented in high--$Q$ cavities a natural alternative is to detect
the photons inside the resonator prior to their escape. In what I call the
\emph{posterior intracavity detection} the detector is inserted inside the
cavity (or turned on) after the time-modulation of the cavity properties
ceased. One of advantages of this approach is that the photon statistics can be
determined at chosen times, but the detector timing must be accurate to avoid disturbing the
generation process and loosing photons due to damping. There are detailed analyzes of
detection of DCE photons in microwave cavities employing antennas \cite%
{padua,braggio}, ensembles of population-inverted alkali-metal atoms \cite{onofrio1,onofrio2},
single Rydberg atoms \cite{yamamoto} and superconducting qubits \cite{zeilinger}. A
proposal based on measuring the change in kinetic energy of electrons
passing through the cavity was given in \cite{electrons}.

In the \emph{continuous intracavity monitoring} the intracavity detector is active from the very beginning of the time-modulation, as
illustrated by the artistic view in figure \ref{f0}. The upside is that
no complex timing techniques are required, however the detector may disturb
significantly the photon generation process or even completely inhibit it if
not properly accounted for. This line started with papers \cite{pla,pra} for
harmonic time-modulation of the cavity frequency, where the detector was modeled
either as a $2$-level atom or a harmonic oscillator. In \cite%
{fed-jetp,fed-pla,fed-pra} non-harmonic perturbations in the presence of a $2$-level atom were analyzed. A simultaneous implementation and detection of a DCE analog
in superconducting circuits using an artificial $2$-level atom with
time-varying transition frequency or coupling parameter was discussed in
\cite{me-arx,jpcs}. Quite recently new results were found for the harmonic
modulation of cavity frequency in the presence of lossy multi-level atoms
\cite{roberto,dodo11,2level,2atom,3level,Nlevel} and harmonic oscillators
\cite{cpl,Sinaia,cacheffo}.

Finally, the method of \emph{back-action detection} aims monitoring the
back-action of the DCE on the external agent responsible for the
time-modulation of the system parameters. For example, the energy of the
photons created due to a moving boundary comes from the mechanical energy,
so the wall experiences a friction force by the quantum vacuum. Reference \cite%
{backaction} analyzed this approach in the case of DCE induced by the Rabi
oscillations of a $3$-level emitter illuminated by a strong laser while coupled to the cavity field mode, and it was shown that a suppression of the
absorption by the emitter can pinpoint the DCE.

In this paper I review the recent results on the continuous intracavity
monitoring of DCE, when the cavity field interacts coherently with a multi-level
\emph{quantum detector} that can
be coupled to an external (classical) measurement device in order to read out its state. It will be shown that the presence of quantum detector
leads to novel regimes of photon generation,
so besides detecting the Casimir photons one could produce novel cavity
field states by parametrically tuning the periodicity of modulation and applying
post-selection techniques. I shall also point out the mathematical
equivalence of DCE in the presence of detectors to the
research in the area of Quantum Nonlinear Optics (QNO), where nonlinear pumped
crystals interact with multi-level atoms and quantum wells.

This paper is organized as follows. The formalism of
continuous intracavity monitoring is outlined in section \ref{CID}, and
the modeling of the detector as $N$-level ladder atom is described in
section \ref{N-level}. The relationship of this topic to the area of QNO is outlined in section \ref{NQO}. The main results and discussion for different values of $N$ and for networks of $2$-level atoms are given in
section \ref{results}. Finally, the sections \ref{future} and \ref{Conclusions} contain some perspectives of future developments and the summary of this brief review.

\section{Continuous intracavity monitoring}

\label{CID}

I shall consider the simplest case of a single resonant cavity mode whose
angular frequency $\omega _{t}$ is rapidly modulated in time around its bare
frequency $\omega _{0}$ according to the harmonical law $\omega _{t}=\omega
_{0}+\varepsilon \sin (\eta t)$ with a small modulation depth $|\varepsilon
|\ll \omega _{0}$. The Hamiltonian describing this cavity DCE implementation
is \cite{Law94} (I set $\hbar =1$)%
\begin{equation}
\hat{H}_{c}=\omega _{t}\hat{n}+i\chi _{t}(\hat{a}^{\dagger 2}-\hat{a}^{2}),
\label{Hc}
\end{equation}%
where $\chi _{t}=(4\omega _{t})^{-1}d\omega _{t}/dt$ is a so-called
squeezing coefficient, $\hat{a}$ and $\hat{a}^{\dagger }$ are the cavity
annihilation and creation operators satisfying the bosonic commutation
relation $[\hat{a},\hat{a}^{\dagger }]=1$, and $\hat{n}=\hat{a}^{\dagger }%
\hat{a}$ is the photon number operator. In the presence of intracavity
quantum detector the total Hamiltonian is $\hat{H}^{\prime }=\hat{H}_{c}+%
\hat{H}_{d}$, where $\hat{H}_{d}$ denotes the detector free Hamiltonian plus
the detector--field interaction. To find the dynamics one has to solve
the von Neumann equation $i\partial \hat{\rho}^{\prime }/\partial t=[\hat{H}%
^{\prime },\hat{\rho}^{\prime }]$ for the total density operator $\hat{\rho}^{\prime }$ of the
detector--field system. Furthermore, due to the
detector--field entanglement the detector may serve to produce novel cavity
field states via post-selection procedures. If the information about the
detector state is completely discarded, then the field state
at time $t$ is simply $\hat{\rho}_{f}^{\prime }(t)=\mathrm{Tr}_{d}[\hat{\rho}%
^{\prime }(t)]$, where $\mathrm{Tr}_{d}$ denotes the partial trace operation
over the detector's degrees of freedom. If the detector state is instead
read in a single-shot projective measurement at the time $t$, with the
outcome described by the projector $\hat{P}_{d}$ on the detector's Hilbert
space, then the cavity field state collapses to the state $\hat{\rho}_{f}^{\prime
}(t_+)={\mathrm{Tr}_{d}[\hat{P}_{d}\hat{\rho}^{\prime }(t)]}/{\mathrm{Tr}[%
\hat{P}_{d}\hat{\rho}^{\prime }(t)]}$, and $\mathrm{Tr}[\hat{P}_{d}\hat{\rho}%
^{\prime }(t)]$ gives the probability for this outcome.

When some external measurement device (EMD) continuously reads out the detector state, the time evolution of the field--detector system
becomes non-unitary as the system is now open. The dynamics can then be described by the Continuous Photodetection Model \cite{SD} or
closely related Quantum Trajectories approach \cite{Carm}. When the EMD
emits a \textquotedblleft click\textquotedblright\ as a result of absorbing one excitation from the detector, it
disturbs the detector--field system by promoting the detector to a lower
energy state. Formally this effect is described by the action of the
\textquotedblleft Quantum Jump superoperator\textquotedblright\ (QJS) $\hat{J%
}$ on the density operator $\hat{\rho}^{\prime }$, and the probability of a
click during the infinitesimal time interval $[t,t+\Delta t)$ is $\,\mathrm{%
Tr}[\hat{J}\hat{\rho}^{\prime }(t)]\Delta t$. The system state immediately
after the click becomes $\hat{\rho}^{\prime }(t_{+})={\hat{J}\hat{\rho}%
^{\prime }(t)}/{\mathrm{Tr}[\hat{J}\hat{\rho}^{\prime }(t)]}$. Between the
clicks the time evolution is also modified due to continuous
leak of information from the system to the EMD, and the dynamics during the time interval $%
[t_{0},t)$ must be described by the non-unitary \textquotedblleft No-count\textquotedblright\ superoperator $\hat{S}%
_{t-t_{0}}\hat{\rho}^{\prime }(t_{0})\equiv \hat{s}_{t-t_{0}}\rho ^{\prime
}(t_{0})\hat{s}_{t-t_{0}}^{\dagger }$. Here $\hat{\rho}^{\prime }(t_{0})$
is the system density operator at the time $t_{0}$ and the operator $\hat{s}%
_{t}$ obeys the differential equation $i\partial \hat{s}_{t}/\partial t=(%
\hat{H}^{\prime }-i\hat{R}/2)\hat{s}_{t}\,$, where $\hat{R}$ is a positive
operator defined by the relation $\mathrm{Tr}[\hat{J}\hat{\rho}^{\prime }]=%
\mathrm{Tr}[\hat{\rho}^{\prime }\hat{R}]$ (for example, if $\hat{J}\hat{\rho}%
^{\prime }=\sum_{i}\hat{L}_{i}\hat{\rho}^{\prime }\hat{L}_{i}^{\dagger }$
then $\hat{R}=\sum_{i}\hat{L}_{i}^{\dagger }\hat{L}_{i}$) \cite{SD}.
Combining the superoperators $\hat{J}$ and $\hat{S}_{t}$ one may evaluate
the system state after an arbitrary sequence of detection events and the
probability of such \textquotedblleft quantum trajectory\textquotedblright. For deeper discussion regarding the DCE under
continuous monitoring by a $2$-level detector see \cite{roberto,dodo11}.

\section{$N$-level ladder atom}

\label{N-level}

In this paper the quantum detector is modeled as a dissipationless $N$-level
atom in the ladder configuration (hence the terms \textquotedblleft atom\textquotedblright\ and \textquotedblleft quantum
detector\textquotedblright\ will be used interchangeably). Assuming that the atom couples to the field via the dipole interaction, the Hamiltonian $\hat{H}_d$ is
\begin{equation}
\hat{H}_{d}=\sum_{j=1}^{N}E_{j}\hat{\sigma}_{j}+\sum_{j=1}^{N-1}g_{j}(\hat{a}%
+\hat{a}^{\dagger })(\hat{\sigma}_{j+1,j}+\hat{\sigma}_{j,j+1}),
\label{totHam}
\end{equation}%
where $E_{j}$ is the energy of the $j$-th atomic state $|\mathbf{j}\rangle $%
, $\hat{\sigma}_{j}\equiv |\mathbf{j}\rangle \langle \mathbf{j}|$ and $\hat{%
\sigma}_{k,j}\equiv |\mathbf{k}\rangle \langle \mathbf{j}|$ are the
generalized Pauli operators and $g_{j}$ (assumed real) is the coupling
parameter between the atomic states \{$|\mathbf{j}\rangle ,|\mathbf{j}+%
\mathbf{1}\rangle $\} via the cavity field. For $|\varepsilon |\ll \omega
_{0}$ the modulation is only relevant for the squeezing coefficient, so one
may write $\omega _{t}\simeq \omega _{0}$ and $\chi _{t}\simeq
\left( \varepsilon \eta /4\omega _{0}\right) \cos (\eta t)$ in the equation (%
\ref{Hc}). For the \textquotedblleft empty cavity\textquotedblright\ (without the detector) the
maximal photon generation occurs for the modulation frequency $\eta =2\omega
_{0}$, so in the following I shall write $\eta =2\omega _{0}+2r$, where $2r$
is the \emph{resonance shift} from the standard DCE\ resonance. In the
interaction picture defined by the time-dependent unitary transformation $%
\hat{\rho}^{\prime }(t)=\hat{V}(t)\hat{\rho}(t)\hat{V}^{\dagger }(t)$ with $%
\hat{V}(t)=\exp [-\frac{1}{2}it(\eta \hat{n}+\sum_{k=0}^{N-1}(2E_{1}+k\eta )%
\hat{\sigma}_{k+1})]$ the rotated statistical operator $\hat{\rho}$ evolves
according to the Hamiltonian $\hat{H}=\hat{V}^{\dagger }\left( t\right) \hat{%
H}^{\prime }\hat{V}\left( t\right) -i\hat{V}^{\dagger }\left( t\right)
\partial \hat{V}\left( t\right) /\partial t$%
\begin{eqnarray}
\hat{H} &=&i\beta _{r}\left( \hat{a}^{\dagger 2}-\hat{a}^{2}\right) -r\hat{n}%
-\sum_{l=1}^{N-1}\sum_{j=1}^{l}(\Delta _{j}+r)\hat{\sigma}_{l+1}  \nonumber
\\
&&+\sum_{l=1}^{N-1}g_{l}(\hat{a}\hat{\sigma}_{l+1,l}+e^{-i\eta t}\hat{a}\hat{%
\sigma}_{l,l+1}+H.c.),  \label{dream}
\end{eqnarray}%
where $\beta _{r}=(1+r/\omega _{0})\varepsilon /4$, $\Delta _{j}=\omega
_{0}-(E_{j+1}-E_{j})$ is the detuning between the cavity bare frequency and
a given atomic transition frequency ($j=1,\dots ,N-1$), and $H.c.$ stands for
the Hermitian conjugate. For $\left\vert g_{j}\right\vert ,|\Delta _{j}|\ll
\omega _{0}$ one can neglect the rapidly oscillating term $(e^{-i\eta t}\hat{%
a}\hat{\sigma}_{i,i+1}+H.c.)$ in the so called Rotating Wave Approximation
(RWA), which we adopt from now on.

In the case of empty cavity, null resonance shift ($r=0$) and initial
vacuum state $|0\rangle_{field}$ \footnote{The field states are expressed in the Fock basis throughout the paper.} the average number of photons increases with time
as $\langle \hat{n}(t)\rangle =\sinh ^{2}(2\beta _{0}t)$, so for large
times $\langle \hat{n}(t)\rangle \simeq \exp (4\beta _{0}t)/4$. The field goes to the squeezed vacuum state, for which only even photon numbers
are present and the field quadrature operators $\hat{x}_{\pm }=(\hat{a}\pm
\hat{a}^{\dagger })/\sqrt{\pm 2}$ have zero averages and variances $\langle
(\Delta \hat{x}_{\pm })^{2}\rangle =\exp (\pm 4\beta _{0}t)/2.$ The
statistics of this state is called sometimes \textquotedblleft
super-chaotic\textquotedblright , because the Mandel factor, $Q\equiv
\lbrack \langle (\Delta \hat{n})^{2}\rangle -\langle \hat{n}\rangle
]/\langle \hat{n}\rangle =1+2\langle \hat{n}(t)\rangle $, is roughly twice
bigger (for $\langle \hat{n}\rangle \gg 1$) than its value $Q_{therm}=\langle \hat{%
n}\rangle $ in the \textquotedblleft usual chaotic\textquotedblright\
thermal state. It will be shown in the next section that even more
\textquotedblleft chaotic\textquotedblright\ states, with $Q>1+2\langle \hat{%
n}\rangle $ (also called \textquotedblleft
hyper-Poissonian\textquotedblright\ \cite{dodo11}), can be produced when DCE
is implemented in the presence of the quantum detector.

\section{Relationship with Quantum Nonlinear Optics}\label{NQO}

The interaction of $2$-level atoms with squeezed states of light has attracted enormous interest in the Quantum Optics community since the prediction by Gardiner \cite{gardiner} in $1986$ that the two polarization components of the atom would be damped at different rates when it interacts with a broadband squeezed vacuum, leading to a longer relaxation time for one of these components when compared to normal vacuum radiative decay. Thereafter a large number of papers addressed different aspects of the problem of atom--field interaction in the presence of squeezing and external driving. In the so called \textquotedblleft passive\textquotedblright\ scheme \cite{49} the atom is driven by a squeezed light produced externally [e.g., by the subthreshold degenerate optical parametric oscillator (DOPO)] or the atom--field system is damped by a broadband squeezed reservoir \cite{JOSAB,42,44,45,46,53,58,59}; this setup parallels the posterior intracavity detection of DCE. Contrary, in the \textquotedblleft active\textquotedblright\ scheme the atom is located inside the squeezing generator (such as DOPO) during the entire operation in order to enhance its coupling to the squeezed field modes and eliminate the propagation losses \cite{40,48,49,opa2,opa3,opa4,opa6,opa8,opa9,opa10,opa12,opa13}; this design formally resembles the continuous intracavity monitoring of DCE.

Different aspects of this subject were investigated over the last $25$ years, including: analytical description of the dynamics and quantum statistical properties of the system \cite{opa7,opa9,opa10,opa3,opa4,opa6,opa8,opa12}, narrowing and hole burning in the spectrum of the fluorescent and transmitted light \cite{opa2,JOSAB,42,45,53,59}, suppression or enhancement of the atomic population decay \cite{42,44}, steady state behavior of atomic population inversion and cavity field correlation functions \cite{40,46,opa9,opa10}, collapse of the atom into a pure state \cite{53}, bistability in the steady-state
intracavity intensity versus pumping intensity \cite{48}, enhancement of the intracavity squeezing in lossy DOPO containing two-level atoms \cite{49}, etc. It is impossible to account for all the pertinent literature in this short article, so I refer the interested reader to the reviews \cite{parkins,dalton} and references in \cite{58,53,opa2}.

Apart from eventual terms describing the external driving and dissipation, the chief Hamiltonian describing the Quantum Nonlinear Optics setups with multi-level atoms or harmonic oscillators (such as other cavity field modes \cite{opa1} or quantum wells \cite{opa13}) is similar to the Hamiltonian (\ref{dream}) describing the cavity DCE in the presence of quantum detectors. However the scope of investigations and the adopted regimes of parameters have been rather different in these areas. In the cavity DCE research the focus has been the dynamics of photon generation from vacuum (and other low-excitation initial states) and the detector--field entanglement under the assumption of small dissipative losses. On the other hand, in the QNO schemes the majority of studies addressed the steady-state behavior, assumed strong damping or external driving of the atom or the cavity, and the dynamics was usually studied for nonvacuum initial states. Hence, these two physically distinct phenomena are complementary from the mathematical standpoint, and future investigations on the continuous intracavity monitoring of the DCE can take advantage of the vast literature concerning the interaction of atoms with light inside optical parametric oscillators.

\section{Analytical and numerical results}

\label{results}

\subsection{Resonant $N$-level atom}

Let us first analyze the detector's influence in the resonant case, when all
atomic transitions are resonant with the field mode: $\Delta _{j}=0$ for $%
j=1,\ldots ,N-1$. The \emph{strong modulation regime} occurs when $\left\vert
\varepsilon \right\vert \gg \left\vert g_{j}\right\vert $; it can be easily
achieved in the circuit QED setup by reducing the atom-field couplings $%
g_{j} $, which currently are $|g_{j}|/\omega _{0}\sim 10^{-2}$ but can be
made small at will \cite{circuitQED}. Performing the time-independent
unitary transformation $\hat{\rho}=\hat{U}\hat{\rho}_{1}\hat{U}^{\dagger }$,
where $\hat{U}=\exp [-i\sum_{l=1}^{N-1}\xi _{l}(\hat{a}\hat{\sigma}%
_{l,l+1}+H.c.)]$ and $\xi _{l}=g_{l}/2\beta _{r}$, the rotated density operator $\hat{\rho}_{1}$ evolves
according to the Hamiltonian $\hat{U}^{\dagger }\hat{H}\hat{U}$. To the
second order in the small parameter of the order $\mathcal{O}(\xi )$
(assuming $\mathcal{O}(\xi _{j})\sim \mathcal{O}\left( \xi \right) $ for all
$j$) for the resonance shift $r=0$ one obtains the effective Hamiltonian%
\begin{equation}
H_{eff}\simeq i\beta _{0}[\hat{\theta}\hat{a}^{\dagger
2}+\sum_{l=1}^{N-2}\xi _{l}\xi _{l+1}\hat{\sigma}_{l,l+2}-H.c.],
\end{equation}%
where $\hat{\theta}=1+\sum\nolimits_{l=1}^{N-1}\xi _{l}^{2}\left( \hat{\sigma%
}_{l+1}-\hat{\sigma}_{l}\right) $. It shows that many photons are generated,
but the photon generation rate is slightly affected by the atom depending on
its initial state; moreover, the atomic levels become coupled via the cavity
field. This Hamiltonian was thoroughly investigated for two-level \cite%
{2level} and three-level \cite{3level} detectors, and exact expressions for
the photon number distribution and other relevant observable quantities were
obtained.

\begin{figure}[th]
\begin{center}
\includegraphics[width=.49\textwidth]{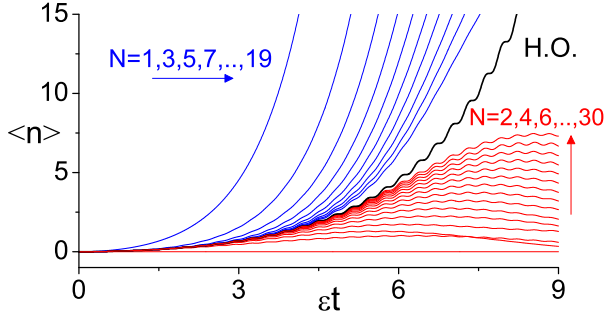}
\end{center}
\caption{Average photon number as function of dimensionless time $\protect%
\varepsilon t$ for different number of levels $N$ and \textquotedblleft harmonic coupling\textquotedblright\ $%
g_{l}=g\sqrt{l}$. H.O. stands for harmonic oscillator. Parameters: $%
\Delta _{l}=0$, $g/\omega_0=10^{-2}$ and $\varepsilon/\omega_0 =10^{-3}$.}
\label{f2}
\end{figure}

The \emph{weak modulation regime}, $\left\vert \varepsilon \right\vert \ll
\left\vert g_{j}\right\vert $, is achieved by increasing the atom-field
couplings beyond the modulation depth; in this case the detector can substantially
modify the photon generation process. In the absence of damping and for
initial state $|\mathbf{1},0\rangle \equiv|\mathbf{1}\rangle_{atom}|0\rangle_{field}$ the system state remains pure
throughout the evolution, so instead of the von Neumann equation in the following I shall consider the Schr\"{o}%
dinger equation $i\partial |\psi \rangle /\partial t=\hat{H}|\psi \rangle $ for the wavefunction $|\psi \rangle$,
where the Hamiltonian $\hat{H}$ is given by the expression (\ref{dream}) under the RWA. For the null resonance shift, if $%
\varepsilon =0$ and the number of levels $N$ is odd, then for $m>0$ excitations the Hamiltonian $%
\hat{H}$ has one null eigenvalue $\lambda _{m,0}=0$ corresponding to the eigenstate $|\phi _{m,0}\rangle =\mathcal{N}_{m}^{-1}%
\sum_{k=0}^{(N-1)/2}\alpha _{k}|\mathbf{2k+1},m-2k\rangle $, where $\mathcal{%
N}_{m}$ is the normalization constant and $\alpha _{k+1}=-\alpha _{k}g_{2k+1}%
\sqrt{m-2k}/(g_{2k+2}\sqrt{m-2k-1})$ \footnote{For $m=0$ the eigenvalue is $0$ and the eigenstate is $|{\bf 1},0\rangle$.}. The other eigenstates with $m$
excitations are denoted by $|\phi _{m,k\neq 0}\rangle $ and the
corresponding eigenvalues $\lambda _{m,k}$ are functions of $g_{j}$ ($%
j=1,\ldots ,N-1$), so they are large compared to $\beta _{0}$. Making the
ansatz $|\psi \rangle =\sum_{m,k}e^{-it\lambda _{m,k}}A_{m,k}(t)|\phi
_{m,k}\rangle $ one arrives at the set of differential equations%
\begin{equation}
\frac{\partial A_{m,j}}{\partial t}=\beta _{0}\sum_{n,k}e^{-it\lambda _{n,k}}\langle \phi _{m,j}|(\hat{a}^{\dagger 2}-\hat{a}%
^{2})|\phi _{n,k}\rangle A_{n,k}
\end{equation}%
with the initial condition $A_{0,0}(0)=1$. Neglecting the rapidly oscillating
terms, for which $|\lambda _{n,k}|\gg |\beta _{0}\langle \phi _{m,j}|(\hat{a}%
^{\dagger 2}-\hat{a}^{2})|\phi _{n,k}\rangle |$, one finds that all the
coefficients $\{A_{m,0},A_{m+2,0}\}$ for $m=0,\ldots ,\infty $ become
resonantly coupled, while the coefficients $A_{m,k\neq 0}$ remain close to
zero throughout the evolution \cite{Nlevel}. Thus the number of created
photons is unlimited and only the states $|\phi _{m,0}\rangle $ become significantly
populated. For an even number $N$ there is a null eigenvalue for the $m$%
-excitation eigenstate $|\phi _{m,0}\rangle $ if $m\leq N-2$, therefore at
most $N-2$ photons can be created in this case \cite{Nlevel}. Such behavior
is shown in figure \ref{f2} for different numbers $N$.

\subsection{Harmonic oscillator}

In the special case when $N\rightarrow \infty $ and $g_{l}=g\sqrt{l}$ the
detector becomes a simple harmonic oscillator (H.O.) if one associates the
atom level $|\mathbf{1}\rangle $ with the oscillator ground (zero energy)
state and makes the replacement $\sum_{l=1}^{\infty }\sqrt{l}\hat{\sigma}%
_{l,l+1}=\hat{b}$, where $\hat{b}$ is the annihilation operator associated
with the detector obeying bosonic commutation relation $[\hat{b},\hat{b}%
^{\dagger }]=1$. Then the Heisenberg equations of motion for operators $\hat{%
a}(t)$ and $\hat{b}(t)$ can be solved exactly \footnote{%
These equations were solved in the context of QNO
in \cite{opa13}, while the solution for $\langle \hat{n}%
\rangle $ was obtained independently in \cite{cpl}.}, and the quadrature
variances become \cite{Nlevel,Sinaia}%
\begin{equation}
\langle (\Delta \hat{x}_{\pm })^{2}\rangle ={e^{\pm 2\beta _{0}t}}%
\left( \frac{1}{2}\pm \frac{\beta _{0}}{2\gamma }\sin (2\gamma t)+\frac{\beta _{0}^{2}%
}{\gamma ^{2}}\sin ^{2}(\gamma t)\right) ~,  \label{pt}
\end{equation}%
where $\gamma =\sqrt{g^{2}-\beta _{0}^{2}}$. For $|g|\gg |\beta _{0}|$ the
rate of photon generation becomes roughly twice smaller than for the empty
cavity and the ratio $\langle \hat{b}^{\dagger }\hat{b}\rangle /\langle \hat{%
a}^{\dagger }\hat{a}\rangle $ is close to unity \cite{Nlevel}. For $\beta _{0}t\gg 1$ the
average photon number $\langle \hat{n}(t)\rangle \approx \langle (\Delta
\hat{x}_{+})^{2}\rangle /2$ increases exponentially at the rate $2\beta _{0}$%
, modulated by some oscillations with the frequency $2\gamma $ (leading to appearance of almost horizontal \textquotedblleft shelves\textquotedblright\ seen in figure \ref%
{f2}), and the Mandel factor is $Q(t)\approx 2\langle \hat{n}(t)\rangle $.
Nonetheless, the state of the field mode is not exactly the vacuum squeezed
one since the uncertainty product $\langle (\Delta \hat{p})^{2}\rangle
\langle (\Delta \hat{x})^{2}\rangle =1/4+(g\beta _{0}/\gamma ^{2})^{2}\sin
^{4}(\gamma t)$ is larger than the minimal possible value $1/4$.

\subsection{Resonances for $r\ne 0$}

Besides the resonance $r=0$, at least two photons can be created from vacuum
for the resonance shift $2r=\pm \sqrt{2g_{1}^{2}+g_{2}^{2}}$, though the
concrete details depend on the concrete form of coupling coefficients $\{g_l\}$. For harmonic oscillator
one recovers the resonances $r=\pm g$, first discovered in \cite{pla} and
analyzed in details in \cite{Sinaia,cacheffo}, for which the average photon number
grows as $\left\langle \hat{n}(t)\right\rangle =\sinh ^{2}(\beta _{0}t)/2$.
For a $3$-level atom \footnote{$3$-level atom also in the
Lambda configuration was analyzed in \cite{3level}.}
the nonzero eigenvalues for $\varepsilon =0$ are $\pm \lambda _{n}$ with eigenvectors $|\phi
_{n,\pm }\rangle =(\sqrt{n}g_{1}|\mathbf{1},n\rangle \pm \lambda _{n}|%
\mathbf{2},n-1\rangle +g_{2}\sqrt{n-1}|\mathbf{3},n-2\rangle )/\sqrt{2}%
\lambda _{n}$, where $\lambda _{n}=\sqrt{ng_{1}^{2}+\left( n-1\right)
g_{2}^{2}}$. Thus for $2r=\pm \lambda _{2}$ there is an oscillation between
the states $|\mathbf{1},0\rangle $ and $|\phi _{2,\pm }\rangle $ with the
angular frequency $\beta _r[1+(g_{2}/2g_{1})^{2}]^{-1/2}$ \cite{3level}%
.

\subsection{Atomic network}

One can easily extend the above results to a network of $(N-1)$
identical two-level atoms with transition frequencies $\Omega $ coupled to
the field mode with the same coupling constant $g$ \cite{Nlevel,2atom}. The
Hamiltonian reads%
\begin{equation}
\hat{H}_{a}=\sum_{j=1}^{N-1}[\Omega \hat{\sigma}_{2}^{\left( j\right)
}+g\left( \hat{a}+\hat{a}^{\dagger }\right) (\hat{\sigma}_{1,2}^{(j)}+\hat{%
\sigma}_{2,1}^{(j)})],  \label{Dicke}
\end{equation}%
where the upper index labels the $j$-th atom. Defining the collective
operators $\hat{S}_{z}=\sum_{j=1}^{N-1}\hat{\sigma}_{2}^{(j)}$, $\hat{S}%
_{+}=\sum_{j=1}^{N-1}\hat{\sigma}_{2,1}^{(j)}$ and $\hat{S}%
_{-}=\sum_{j=1}^{N-1}\hat{\sigma}_{1,2}^{(j)}$ the Hamiltonian (\ref{Dicke})
becomes $\hat{H}_{a}=\Omega \hat{S}_{z}+g(\hat{a}+\hat{a}^{\dagger })(\hat{S}%
_{+}+\hat{S}_{-})$. For identical atoms one can introduce the normalized
Dicke state $|\mathbf{j}%
\rangle =\sqrt{[\left( N-1\right) !]^{-1}\left( j-1\right) !\left(
N-j\right) !}\sum_{p}|\mathbf{2}^{\left( 1\right) }\rangle |\mathbf{2}%
^{\left( 2\right) }\rangle \cdots |\mathbf{2}^{\left( j-1\right) }\rangle \\ |%
\mathbf{1}^{\left( j\right) }\rangle \cdots |\mathbf{1}^{\left( N-1\right)
}\rangle $ with $(j-1)$ excitations, where the sum runs over all the allowed permutations of excited
and non-excited atoms and $|\mathbf{k}^{(j)}\rangle $ denotes the state ($k=1,2$) of the $j$-th atom. Using the known properties $\hat{S}%
_{-}|\mathbf{j}\rangle =\sqrt{\left( j-1\right) \left( N-j+1\right) }|%
\mathbf{j-1}\rangle $, $\hat{S}_{+}|\mathbf{j}\rangle =\sqrt{j\left(
N-j\right) }|\mathbf{j+1}\rangle $ and $\hat{S}_{z}|\mathbf{j}\rangle
=\left( j-1\right) |\mathbf{j}\rangle $, in the Dicke basis the Hamiltonian (%
\ref{Dicke}) becomes identical to the Hamiltonian (\ref{totHam}) if one
identifies $E_{j}=\Omega \left( j-1\right)$ and $g_{j}=g\sqrt{j\left( N-j\right)
}$. Thus the network of $(N-1)$ identical two-level atoms is equivalent to an
equidistant $N$-level ladder atom.

\begin{figure}[th]
\begin{center}
\includegraphics[width=.49\textwidth]{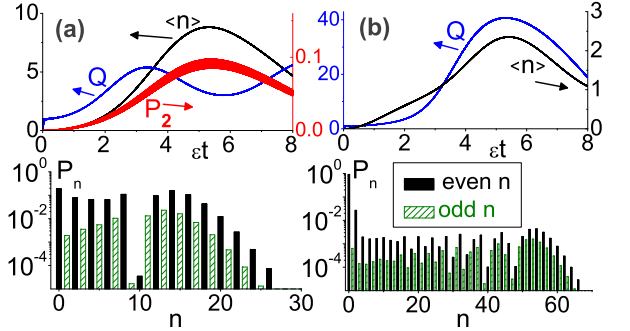}
\end{center}
\caption{Dynamical behavior for $2$-level detector and parameters $\Delta
_{1}=8g_{1}$, $g_{1}/\protect\omega _{0}=10^{-2}$, $\protect\varepsilon /%
\protect\omega _{0}=3\times 10^{-4}$. Correctional shifts are: a) $y=-\varepsilon/2$ and b) $y=-2\varepsilon$. In the bottom are shown the
photon statistics for $\varepsilon t=5$.}
\label{f3}
\end{figure}

\subsection{Two-level atom}

Finally I outline the case of a $2$-level quantum detector. Performing the
unitary transformation $\hat{V}_{2}(t)=\exp [irt\left( \hat{n}+\hat{\sigma}%
_{2}\right) ]$ on the equation (\ref{dream}) the new Hamiltonian reads (under the RWA)%
\begin{equation}
\hat{H}_{2}=-\Delta _{1}\hat{\sigma}_{2}+(i\beta _{r}\hat{a}^{\dagger
2}e^{-2irt}+g_{1}\hat{a}\hat{\sigma}_{2,1}+H.c.)~.
\end{equation}%
Without the modulation, $\varepsilon =0$, this is the celebrated
Jaynes-Cummings model, whose ground state is $|\phi _{0}\rangle =|\mathbf{1}%
,0\rangle $ with eigenenergy $\lambda _{0}=0$, and the excited eigenstates
with $n>0$ excitations are $|\phi _{n,+}\rangle =\sin \theta _{n}|\mathbf{1}%
,n\rangle +\cos \theta _{n}|\mathbf{2},n-1\rangle $, $|\phi _{n,-}\rangle
=\cos \theta _{n}|\mathbf{1},n\rangle -\sin \theta _{n}|\mathbf{2}%
,n-1\rangle $ with eigenenergies $\lambda _{n,\pm }=-\Delta _{1}/2\pm z_{n}$%
, where $z_{n}=\sqrt{( \Delta _{1}/2) ^{2}+g_{1}^{2}n}$ and $%
\theta _{n}=\arctan \sqrt{(z_{n}+\Delta _{1}/2)/(z_{n}-\Delta _{1}/2)}$.

Writing the ansatz%
\begin{equation}
|\psi _{2}(t)\rangle =A_{0}(t)|\phi _{0}\rangle +\sum_{n>0} \sum_{\mathcal{S}=\pm
}e^{-it\lambda _{n,\mathcal{S} }}A_{n,\mathcal{S} }(t)|\phi _{n,\mathcal{S} }\rangle
\end{equation}%
one can check that for $\left\vert g_{1}\right\vert \gg \left\vert \varepsilon
\right\vert $ the coefficient $A_{0}$ couples resonantly to $\{A_{2n,\mathcal{S}}\}$ for the resonance shift $2r=\mathcal{S}z_{2}-\Delta _{1}/2+y$ (where $%
\mathcal{S}=\pm $ and $y$ is a small correctional shift of the order of $\varepsilon$), where $n=1$ if $\Delta
/\left\vert \Delta \right\vert =-\mathcal{S}$ and $n\geq 1$ if $\Delta
/\left\vert \Delta \right\vert =\mathcal{S}$. In the resonant regime, $|\Delta _{1}|\ll |g_{1}|$,
at most two photons can be generated \cite{pla,roberto}. On the other hand, in the dispersive
regime, when $(\Delta _{1}/2)^{2}\gg g_{1}^{2}n$ for all relevant $n$, one has $%
z_{n>0}\simeq \left\vert \Delta _{1}\right\vert /2+\left\vert \delta
\right\vert n$ and $\lambda _{n+2,\pm }-\lambda _{n,\pm }=$ $\pm 2\left\vert
\delta \right\vert $, where $\delta =g_{1}^{2}/\Delta _{1}$ is the cavity
dispersive shift due to the detector; hence many photons can be created from
the initial state $|\mathbf{1},0\rangle $ for the resonance shift $2r=2\delta $
\cite{2atom}.

The most interesting case occurs when $|\Delta
_{1}|\sim |g_{1}|$, as in this case the detector acts as saturable limiter
and exotic field states can be produced \cite{me-arx,jpcs,roberto,dodo11}.
In figure \ref{f3} are plotted $\left\langle \hat{n}\right\rangle $, $Q$%
, the atomic excitation probability $P_{\bf 2}$ and the photon number
distribution at some time instant for two different values of $y$.
One can see that the system dynamics is very sensitive to small changes in
the modulation frequency, and the \textquotedblleft hyper-Poissonian\textquotedblright\ states appear quite
naturally (figure \ref{f3}b). Detailed studies of this regime for different
initial states and in the presence of EMD can be found in \cite%
{roberto,dodo11}.

\section{Discussion and future directions}\label{future}

An important actual task is to study how the detector's temperature and
damping influence the photon generation regimes under the continuous intracavity monitoring (the cavity damping was considered in numerous papers, see \cite{vdodonov} for a brief review). Preliminary numerical results for the case when the detector can be modeled as a $2$-level atom were given in \cite{roberto}, while the paper \cite{3level} analyzed numerically the situation for $3$-level atoms in the ladder or $V$ configurations. Owing to dissipation the dynamics of photon generation from vacuum is qualitatively different from the lossless case, as system states that could not be populated via the unitary evolution now can become populated via dissipative channels. Moreover, for weak dissipation, the photon number probabilities exhibit oscillatory behavior as function of time before attaining the steady-state values \cite{3level}, so both the dynamics and the stationary properties deserve further investigations. In this connection the rich literature on the steady-state behavior of atoms inside squeezing generators within the QNO area may be useful to foresee the stationary features in the realistic continuous intracavity monitoring of DCE.

The modeling of detector as harmonic oscillator also seems to be rather realistic in the so called \textquotedblleft Motion Induced Radiation\textquotedblright\ (MIR) experiment \cite{padua,braggio}, where the microwave quanta created via DCE are supposed to be detected by means of a small antenna put inside the cavity.
Since the inductive antenna (a wire loop) used in that experiment is a part of a LC-contour,
it can be reasonably approximated as a harmonic oscillator. To our knowledge the effects of dissipation on the photon production via DCE have not been addressed for this type of detector, although a thorough mathematical analysis of a similar problem was carried out within the scope of QNO \cite{opa13} and the results could be easily transposed to the DCE scenario.

Another interesting perspective is to exploit the rich
eigenvalue spectrum of the atom--field interaction Hamiltonian to implement
multi-modulation regimes, when the modulation consists of multiple sinusoids
whose frequencies may adiabatically vary with time \cite{jpcs}. In such a way one could selectively couple different sets
of the dressed states of the atom--field system and control the amount of photons generated from vacuum. Finally, the implementation of analogs of DCE in highly controllable solid state systems may lead to novel schemes of generating nonclassical states of light and manipulating the light--matter interaction.

\section{Summary}

\label{Conclusions}

I summarized the main results on the continuous intracavity monitoring
of the dynamical Casimir effect using multi-level quantum detectors that may
be coupled to classical external measurement devices. It was shown
that the photon generation from vacuum can be severely
affected by the detector, therefore the periodicity of external modulation
must be parametrically tuned to induce different dynamical regimes. In the absence of dissipation and for specific modulation frequencies the unitary evolution from vacuum can lead to field--detector entangled states with: (i) at most two photons, (ii) a small amount of photons ($>2$) or (iii) unlimited
photon number. Thus at the expense of reducing the degree of squeezing and
the number of created photons (or the rate of exponential photon growth), the photon generation from vacuum can be continuously monitored, and qualitatively new
quantum states could be generated using the post-selection procedures.

\begin{acknowledgments}
The present work was supported by CNPq, Conselho Nacional de Desenvolvimento Científico e Tecnológico -- Brazil. Partial support by Decanato de Pesquisa e P\'{o}s-Gradua\c{c}\~{a}o (DPP--UnB, Brazil) is acknowledged.
\end{acknowledgments}

\end{document}